\documentclass[11pt,twoside]{article}

\usepackage{asp2004}
\usepackage{epsf}
\usepackage{epsfig}
\usepackage{lscape}
\usepackage{graphicx}

\begin{document}
\title{Infrared Constraints on AGN Tori Models}   
\author{Hatziminaoglou E. (1), Fritz J. (2), P\'erez-Fournon I. (1), Franceschini A. (2),
Hern\'an-Caballero A. (1), Afonso-Luis A. (1), Lonsdale C. (3), Fang F. (3), Oliver S. (4), 
Rowan-Robinson M. (5), Shupe D. (3), Smith H. (6), Surace J. (3), Gonz\'alez-Solares E. (7), 
and the SWIRE Team}   
\affil{(1) Instituto de Astrofisica de Canarias, C/ Via Lactea s/n, 38200 La Laguna, Spain;
(2) Dipartimento di Astronomia, Universita di Padova, Vicolo Osservatorio 5, 35122 Padua, Italy;
(3) Infrared Processing and Analysis Center, California Institute of Technology, Pasadena, CA 91125, USA;
(4) Astronomy Centre, Department of Physics and Astronomy, University of Sussex, Falmer,
Brighton BN1 9QJ, UK;
(5) Astrophysics Group, Blackett Laboratory, Imperial College London, London SW7 2BW, UK;
(6) Center for Astrophysics and Space Sciences, University of California, San Diego, La Jolla,
CA 92093-0424, USA;
(7) Institute of Astronomy, University of Cambridge, Madingley Road, Cambridge CB3 0HA, UK}

\begin{abstract} 
This work focuses on the properties of dusty tori in active galactic nuclei (AGN) derived
from the comparison of SDSS type 1 quasars with mid-Infrared (MIR) counterparts
and a new, detailed torus model. The infrared data were taken by the Spitzer Wide-area
InfraRed Extragalactic (SWIRE) Survey. Basic model parameters are constraint,
such as the density law of the graphite and silicate grains, the torus size and its
opening angle. A whole variety of optical depths is supported.  The favoured models
are those with decreasing density with distance from the centre, while there is no
clear tendency as to the covering factor, ie small, medium and large covering factors are
almost equally distributed. Based on the models that better describe the observed SEDs,
properties such as the accretion luminosity, the mass of dust, the inner to outer radius ratio
and the hydrogen column density are computed.
\end{abstract}

\section{Torus Model}   

The torus geometry adopted to describe the shape and the spatial distribution of dust 
is the so-called {\it flared disk}, that is a sphere with the polar cones removed. Its size 
is defined by its outer radius, R$_{\rm out}$, and the opening angle, of the torus itself. The dust
components that dominate both the absorption and the emission of radiation are
graphite and silicate. The location of the inner radius, R$_{\rm in}$, depends both on the
sublimation temperature of the dust grains (1500 and 1000 K, for graphite and silicate,
respectively) and on the strength of the accretion luminosity. We adopted the absorption
and scattering coefficients given by \cite{laor93} for dust grains of different
dimensions, weighted with the standard MRN distribution \citep{mathis77}.
Grains dimensions range from 0.005 to 0.25 micron for graphite, and 0.025 to 0.25
micron for silicate. The gas density within the torus is modeled in such a way to allow
a gradient along both the radial and the angular coordinates.

The central source is assumed to be point-like and its emission isotropic. Its spectral
energy distribution is defined by means of a composition of power laws with
different values for the spectra index in the UV, optical and IR. A numerical approach 
-- the $\Lambda$-iteration method -- was adopted to solve the radiative transfer equation.
A geometrical grid is defined along the three spatial coordinates, and the main
physical quantities (dust density and temperatures, electromagnetic emission, optical
depth, etc.) are computed with respect to the center of the volume elements defined
by the grid. When computing the total incoming energy on a given volume
element a very accurate computation of the optical depth, is done. Furthermore,
since silicate grains have a lower sublimation temperature with respect to graphite,
we account for the fact that the innermost regions of the torus contain only graphite 
and are silicate free
 
\begin{figure}[!ht]
\includegraphics[angle=-90, width=12cm]{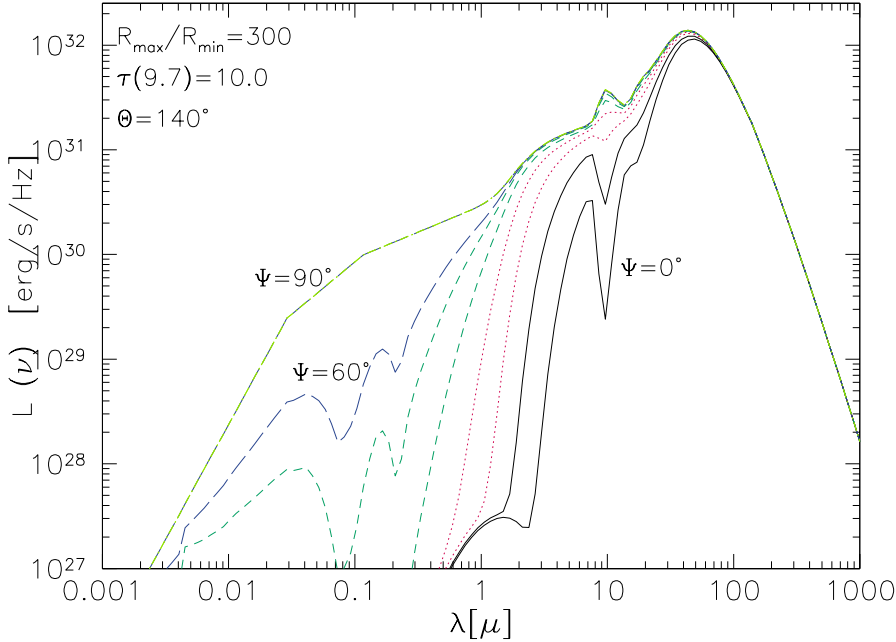}
\caption{Emission spectra as a function of wavelength for 10 different
lines-of-sight inclinations from ${\Psi} = 0^\circ$ (edge-on; lower curve) to ${\Psi} = 90^\circ$ 
(face-on, upper curve) at regular steps of $10^\circ$, for a geometrical
configuration with $\tau(9.7)=10.0$, R$_{\rm out}$/R$_{\rm in}=300$, torus
opening angle $\Theta=140^\circ$ and a density that depends on both the angle and the distance
from the centre, like 
$\rho\left(r,\theta \right) \propto r^{-1} e^{-6 |cos(\theta)|}$}
\label{figMod}
\end{figure}

Finally, the global SED is computed at different angles of the line-of-sight with respect
to the torus equatorial plane, in order to account for both type 1 and type 2 objects
emission and includes three contributions: emission from the AGN, thermal emission
and scattering emission by dust in each volume element. 
For an example of an emitted spectrum see Fig. \ref{figMod}. 
We refer to \cite{fritz06} for
a detailed description of the model developed to compute the emission of dust in AGN.

\section{Data}

The MIR data used here are taken from the SWIRE ELAIS N1 and N2 and the
Lockman fields, and were obtained from February 2004 through July 2004, with
both IRAC and MIPS. For the purposes of this work only the four IRAC bands
and MIPS 24 micron  are used. The SWIRE catalogues we use were processed by the
SWIRE collaboration. Details about the data can be found in \cite{lonsdale04},
\cite{surace05} and Shupe et al., in prep.

The Sloan Digital Sky Survey has validated and made publicly available
its Data Release 4 (DR4), covering the entire Lockman and SWIRE EN2
fields and a part of the SWIRE EN1 field. A total of 280 spectroscopically
confirmed quasars lie within the SWIRE fields covered by the SDSS DR4
spectroscopic release. 
Their $i$-band magnitudes reach 19.1 for objects with redshifts
typically less than 2.3 and go up to a magnitude deeper for higher
redshifts \citep{richards02}. 
For a detailed analysis of the properties of 
the sample in EN1 see \cite{eva05}

\section{First Results}

\begin{figure}[!ht]
\plotone{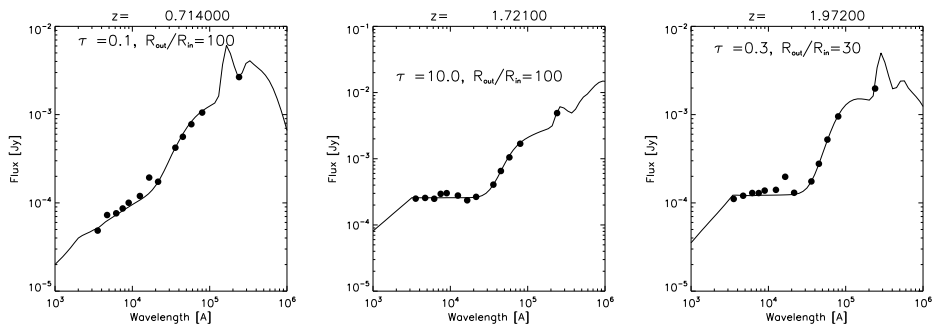}
\caption{Examples of SED fits for three AGN of our sample. The fluxes
are given in Jy, the wavelengths in \AA.}
\label{figEx}
\end{figure}

Each observed SED is compared to a total of 720 models, with 10 lines of
sight each (from an equatorial to a pole-on view). Examples of best 
fits are shown in Fig. \ref{figEx}. Here we summarize
some of the results of the fitting procedure:
\begin{itemize}
\item[$\bullet$] {\it Size of Torus:} Overall, R$_{\rm out}$/R$_{\rm in}$ and the torus 
covering factor are equally distributed between the
allowed values (30, 100, and 300; 95\%, 75\% and 50\%, respectively). However, both
inner and out radii clearly increase with $z$, whereas larger covering factors models
tend to fit realtively low redshift ($z<$ 1) objects.
\item[$\bullet$] {\it Density:}
For $\sim$70\% of the objects the density 
(written as $\rho\left(r,\theta \right) \propto r^{\beta} e^{-\gamma |cos(\theta)|}$) 
decreases radially from the centre while
for another $\sim$25\% it remains constant. 
\item[$\bullet$] {\it Optical depth:}
This work addresses, among other things, the possibility of low optical depth
tori. The SED fitting resulted in some $\sim$45\% of objects with $\tau_{9.7} >$ 1.0 and
$\sim$55\% with $\tau_{9.7} <$ 1.0. This implies that, for some configurations,
type 1 AGN could be seen even when the line of sight intercepts the torus.
\end{itemize}

\begin{figure} 
\plottwo{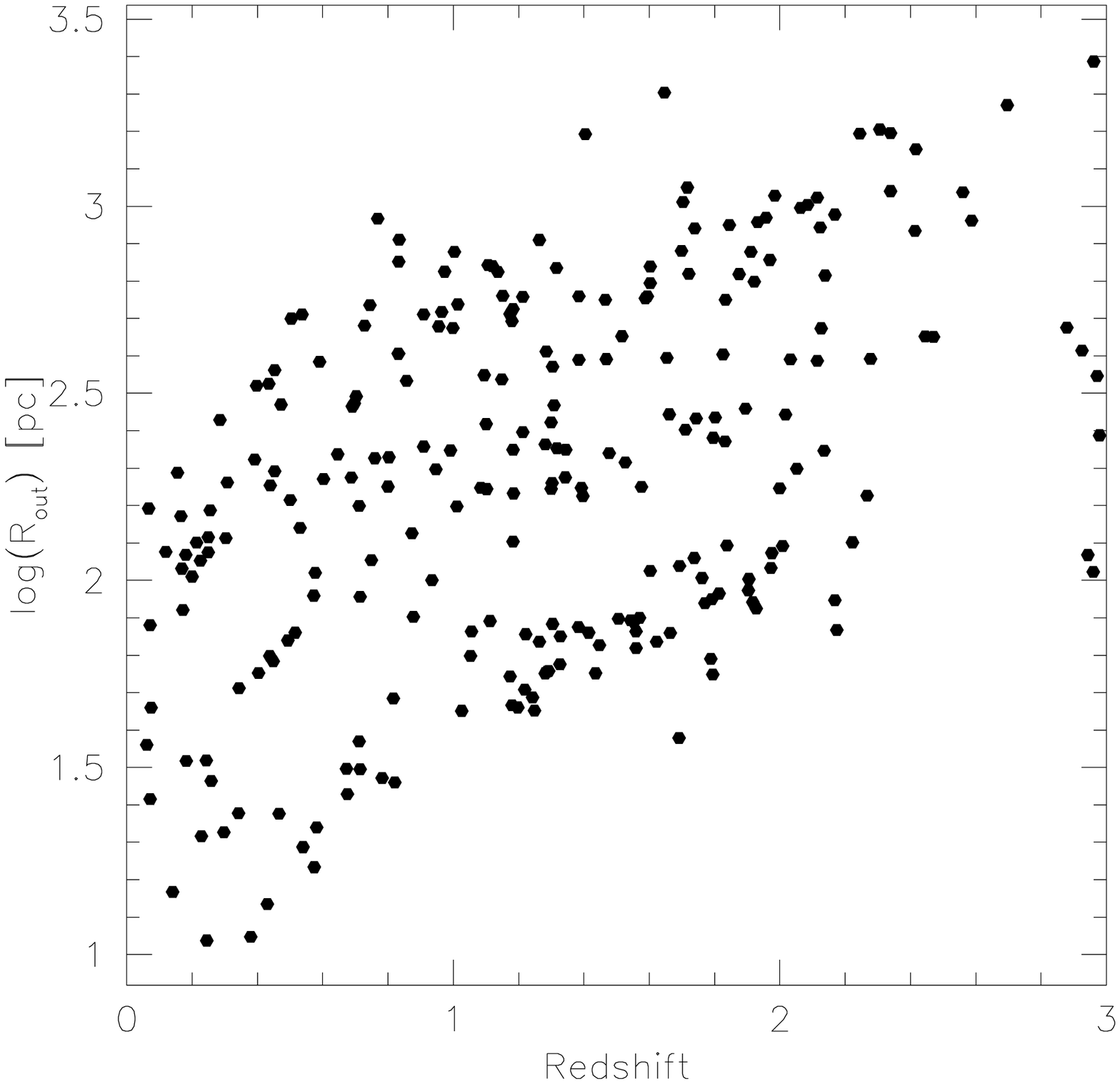}{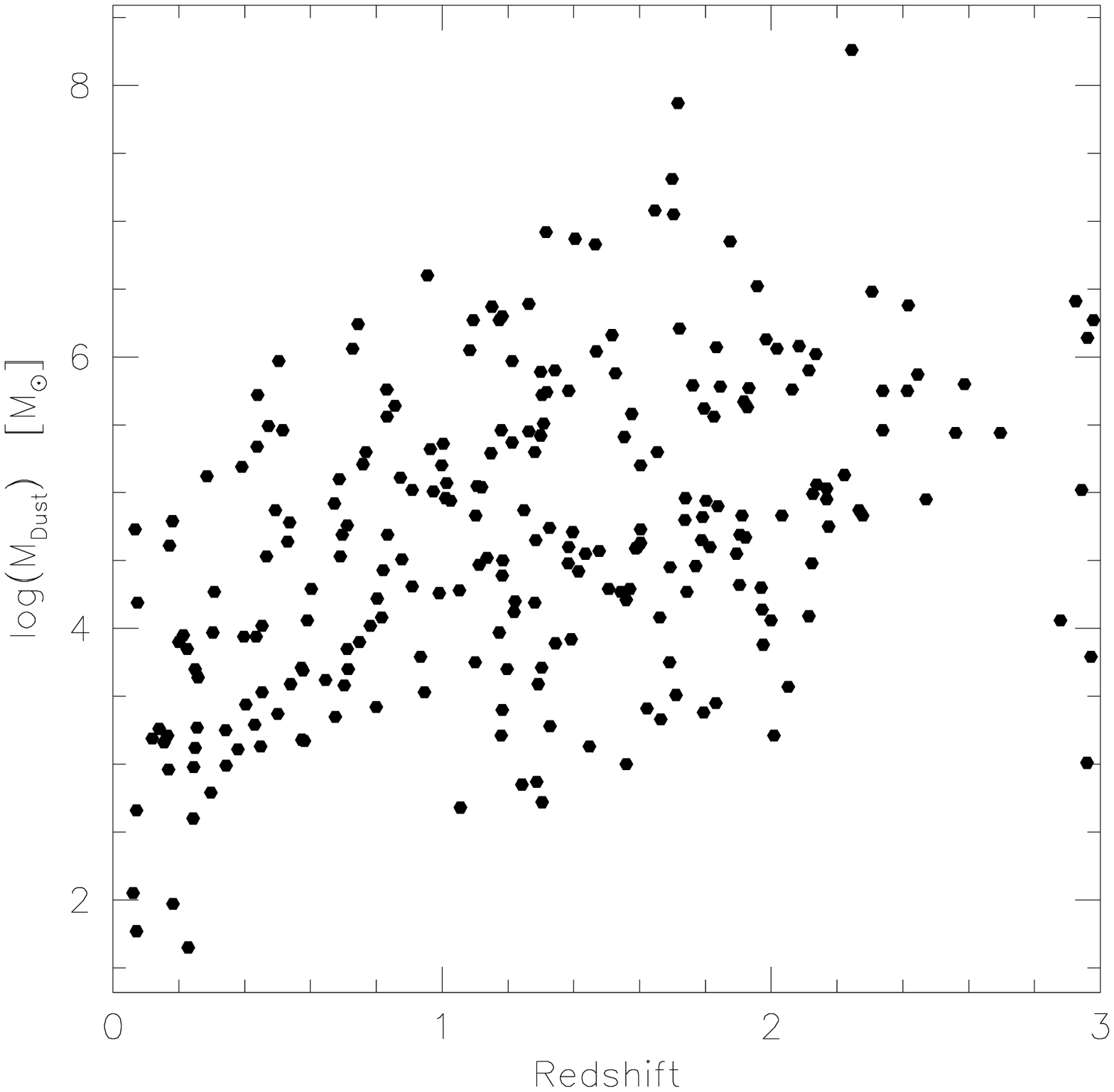}
\caption{Left panel: Outer torus radius, R$_{\rm out}$, as a function of redshift, $z$,
showing a tendency of larger tori at higher $z$.
Right panel: Mass of dust, M$_{\rm Dust}$, as a function of $z$, indicating more massive tori at higher
redshifts.}
\label{figRMz}
\end{figure}

\noindent
The distribution of outer radius of the torus, R$_{\rm out}$, and the mass of dust, M$_{\rm Dust}$, 
with $z$ is shown in the left and right panels of Fig. \ref{figRMz}, respectively.
M$_{\rm Dust}$ is computed summing the individual sample elements of the best-fit model.
The increase of both R$_{\rm out}$ and M$_{\rm Dust}$ with redshift could indicate larger
and more massive tori at higher redshifts, however it could also simply be the manifestation
of the Malmquist bias, as both quantities scale with the accretion luminosity, L$_{acc}$,
an input of our model that is proportional to the bolometric luminosity computed from the data. 
Note that M$_{\rm Dust}$ does not refer to the total mass of the torus, 
that can be obtained by adding the mass of gas, typically $\sim$100 times larger than that of 
the dust.

\acknowledgements 
This work is based on observations made with the {\it Spitzer Space Telescope}.
It also makes use of the SDSS Archive. It was supported in part by the Spanish Ministerio de
Ciencia y Tecnologia (Grants Nr. PB1998-0409-C02-01 and ESP2002-03716)
and by the EC network "POE'' (Grant Nr. HPRN-CT-2000-00138).

\end{document}